# A Deep-learning Approach for Prognosis of Age-Related Macular Degeneration Disease using SD-OCT Imaging Biomarkers

Imon Banerjee[1,†], Luis de Sisternes,[2,†], Joelle Hallak[3], Theodore Leng[4], Aaron Osborne[5], Mary Durbin[2], Daniel Rubin,[1]

[1]Department of Biomedical Data Science, Stanford University, Stanford, CA 94305, USA.
[2] Carl Zeiss Meditec, Inc., Dublin, CA 94568, USA
[3] Department of Ophthalmology and Visual Sciences, University of Illinois at Chicago, IL 60612, USA
[4] Byers Eye Institute at Stanford, Stanford University School of Medicine, Palo Alto, CA 94303, USA
[5] Genentech, CA 94080, USA
*imonb@stanford.edu
[†]Equal contribution

**Abstract** We propose a hybrid sequential deep learning model to predict the risk of AMD progression in non-exudative AMD eyes at multiple timepoints, starting from short-term progression (3-months) up to long-term progression (21-months). Proposed model combines radiomics and deep learning to handle challenges related to imperfect ratio of OCT scan dimension and training cohort size. We considered a retrospective clinical trial dataset that includes 671 fellow eyes with 13,954 dry AMD observations for training and validating the machine learning models on a 10-fold cross validation setting. The proposed RNN model achieved high accuracy (0.96 AUCROC) for the prediction of both short term and long-term AMD progression, and outperformed the traditional random forest model trained. High accuracy achieved by the RNN establishes the ability to identify AMD patients at risk of progressing to advanced AMD at an early stage which could have a high clinical impact as it allows for optimal clinical follow-up, with more frequent screening and potential earlier treatment for those patients at high risk.

## Introduction

Age-related macular degeneration (AMD) is the leading cause of visual loss in developed countries with an aging population. It can be characterized by few symptoms in the early stages but can progress to sudden and severe visual impairment in a late stage[1]. The worldwide prevalence of early stages of AMD in patients between 45 and 85 years is 8% (95% credible interval [3.98–15.49]) and of late AMD is 0.4% (95% credible interval 0.18–0.77)[2]. Prevalence increases with age, reaching 30% for early and 7% for late AMD among those aged 85 years old. Given the increase in life expectancy, nearly 200 million people are expected to have AMD by 2020 and 288 million by 2040. AMD stages can be mainly characterized as "dry" and "wet". Dry AMD is a non-exudative stage with a slow progressive dysfunction of the retinal pigment epithelium (RPE), photoreceptor loss, and retinal degeneration. Early AMD always appears in the dry form characterized by drusen, accumulations of extracellular material that build up between Bruch's membrane and the RPE. This early manifestation can suddenly progress to a non-exudative advanced stage by the appearance of geographic atrophy (GA) or to wet AMD, an exudative stage characterized by the growth of abnormal blood vessels underneath the retina (choroidal neovascularization, CNV) that can leak fluid and blood, leading to swelling and damage of the macula. For simplicity, this article refers to "AMD progression" as the first occurrence of a CNV or an exudative event, where a dry AMD patient converts to a wet stage. Eyes can convert from dry to wet AMD suddenly, where 90% of patients whose disease progresses to wet AMD may lose vision[3]. While no proven treatment exists for advanced AMD in its dry form, early therapeutic intervention with injections is imperative for wet AMD, which makes the effective identification of possible AMD progression highly dependent on regular monitoring through imaging and vision screening. Early detection and intervention in advanced wet AMD have been shown to improve visual outcomes. Moreover, identifying the risk of AMD progression at a given time would allow for optimal clinical follow-up, with more frequent screening, and potential earlier treatment, leading to better clinical outcomes in high-risk patients.

Recent machine learning studies[4], [5] utilized genetic information to predict the risk of AMD with high accuracy, showing an area under a receiver operating characteristic curve (AUC) of around 0.8. Studies[6] aiming to predict the progression to advanced AMD integrating genetic and clinical data even led to AUCs of around 0.9, showing that the combination of genetic and clinical variants are highly significant for the prediction of AMD progression. However, these models mainly showed success in predicting long-term risk AMD progression (>5-years). Additionally, differences in population genetics may limit the application of current prognostic genetic tests as the majority of AMD genetic associations so far have been studied mainly in populations of European ancestry. These findings suggest the need of additional gene variant studies in AMD extending to different populations. In addition to genetic profiling, appropriate individual demographic data is required if individual risk profiles are to be appropriately generated.

Different imaging modalities produce high resolution retinal images disclosing phenotypes that may be good alternatives as non-genomic biomarkers for AMD progression and therapy response. Optical coherence tomography (OCT) is an in vivo imaging method capable of resolving cross-sectional retinal substructures, and the spectral-domain OCT (SD-OCT) represents a gold standard in diagnostic imaging and management of macular diseases given its very fast scanning (more than 20,000 axial scans per second) over a retinal area, with axial resolutions as low as 5 µm. The high acquisition speed and sensitivity of SD-OCT allows the collection of high-resolution three-dimensional (3D) scans while minimizing artifacts due to patient movement or ocular contractions. Previous studies indicate that a degenerative retinal process is associated with the volume of drusen observed in SD-OCT imaging[7], and many other characteristics quantified via SD-OCT may be useful as disease biomarkers. Specific methods of artificial intelligence (machine learning and deep learning) are being increasingly used for automated image analyses in SD-OCT[8]. For instance, traditional machine learning techniques[9]–[13] were used to quantify of specific disease features, and these techniques could also be used for identification of hidden patterns that can be used to improve the prognosis or response to treatment more efficiently. Studies using deep learning[14], [15] have shown to be effective for classifying normal versus AMD OCT images by directly analyzing the image pixel data. Regarding the prediction of AMD progression based on OCT image biomarkers, two main studies have been published previously[11], [16]. These studies propose an initial solution for the prediction of wet AMD using traditional machine learning techniques with limited accuracy (0.74 and 0.68 AUC). Yet, none of the predictive models considered sequential learning on the longitudinal OCT data captured during multiple visits. The predictive performance could be improved by using a sequential deep learning model that considers multiple visits of the same patient in an end-to-end model to predict patient-specific trends in short- and long-term progression of AMD.

However, a practical challenge for implementing such a sequential deep learning model for processing longitudinal raw SD-OCT images is that a single SD-OCT volume typically contains from 100 to 200 two-dimensional (2D) high-resolution images (B-scans), which makes the input data dimension extremely large ($number\ of\ visits \times number\ of\ B-scan\ images \times number\ of\ pixel\ in\ each\ image$) to be handled in a computationally efficient way. A deep learning model that can handle such a complex data space needs huge amount of training data to suppress optimistically biased evaluations of the performance. As a general rule of thumb, the size of dataset should be at-least about 10x its dimension, which is impractical for most of the longitudinal clinical prediction cases, given the limitation of data availability.

To overcome these challenges, we proposed a hybrid modeling approach, which integrates radiomics image features [17] and deep learning in the same platform by extracting the radiomics features from the longitudinal OCT scans and feeding them into a sequential deep learning model for temporal prediction of AMD progression. We analyzed a longitudinal OCT scan dataset of a total of 671 fellow eyes with 13,954 non-neovascular observations using both traditional machine learning and a sequential deep learning technique. We aimed to predict AMD progression at varying time frames, starting from 3 months (short-term) up to 21 months (long-term). Our study provides an interesting insight about the comparative performance of deep learning and traditional machine learning approaches for sequential AMD progression prediction (see *Overall performance*). We also compare the models in a visit variant setting that represents the model performance at a very early point of the sequence and also at a late point, when the model has already learned the trend of the patient by analyzing the previous visit sequence (see *Visit-variant performance*).

**Results**

*Study Dataset*

This study was a retrospective, post hoc analysis of the fellow eye of participants from the HARBOR trial (ClinicalTrials.gov identifier: NCT00891735) [18], a 24-month, phase III study conducted at 100 investigator sites that evaluated the efficacy and safety of 2 doses and 2 regimens of ranibizumab in 1097 patients aged ≥50 years. Patients in this study were selected to have CNV lesions with classic CNV component or occult CNV in the study eye at baseline. All patients had monthly evaluations of both study and fellow eyes with SD-OCT imaging following a standardized protocol acquired with a Cirrus HD-OCT device (Carl Zeiss Meditec, Inc., Dublin, CA, USA) using one of the two following macular scanning protocols, both covering a volume of 6 × 6 × 2 mm$^3$: 512 × 128 × 1024 voxels with a size of 11.7 × 47.2 × 2.0 μm$^3$; or 200 × 200 × 1024 voxels with a size of 30.0 × 30.0 × 2.0 μm$^3$.

Out of the 1097 patients considered in the HARBOR study 686 fellow eyes were graded as non-neovascular at baseline and were considered in this work. Fellow eyes were determined to be non-neovascular at baseline if no neovascular AMD was reported from the eye history case report forms and if no CNV was present at baseline. From this set of fellow eyes, a number of the monthly OCT scans were excluded due to complications linking the OCT data to the clinical diagnosis data resulting from the anonymization process. No quality measures were further considered to exclude any of the remaining images.

The remaining imaging data considered consisted on 686 fellow eyes with a total of 32670 OCT scans. Multiple fellow eyes had several OCT scans acquired during the same imaging session (same date), often acquired using a combination of the two described scan patterns. Imaging features (quantification of any of the properties described in *Materials and Methods*) extracted from OCT scans acquired during the same imaging session where averaged by taking their mean value and considered as a unique observation. OCT scans after a CNV or exudation event (AMD progression) were excluded from the analysis since this work aims to identify features as biomarkers before such events. If no OCT scan from a fellow eye was retrieved before an eventual event, the fellow eye was dropped from the study. AMD progression was determined if a fellow eye that was non-neovascular at baseline received any treatment for exudative AMD at any point during the 2year follow-up. This study then considered a total of 671 fellow eyes with 13,954 non-neovascular observations at unique times prior to progression. Of the 671 fellow eyes, 149 eyes presented progression before the end of the study, while the remaining 522 did not.

*Outcome labeling*

The observations collected from the OCT data were labeled according to the occurrence of a progression event within a given time frame or the certainty of no event within that time frame. Considering a given time, an observation was labeled with a positive outcome ("progressor") if treatment was recorded for that particular eye within that given time elapsed from the observation. On the other hand, an observation was labeled with a negative outcome ("non-progressor") if treatment was not received for that particular eye within the given time frame and there was a subsequent observation of the same eye after the given time where there was no treatment received (no evidence of CNV or exudation). If an observation had no history of treatment within the given time frame and there was no evidence of lack of CNV or exudation after the time frame (either because the end of the study time or missing data), the observation was dropped from the study for that analyzed time, since there were no guarantees that the eye did not progress within that time.

Seven different time frames were analyzed in this study, ranging from 3 months to 21 months within the observation time in 3-month increment intervals. The distribution of number of observations at each given category (progressors and non-progressors) is shown in Fig. 1. As shown, the number of total observations as well as the number of non-progressor observations reduces for longer time intervals due to the length of the study, where observations acquired in the later months of the study duration had no future information available in the longer time intervals (for example, an observation collected at month 12 of the study from an eye that did not develop exudation during the study had future information within 3 months, but not within 21 months due to the 24-month total study time). It can also be seen that the number of progressor observations increases for longer time intervals, as observations that progressed within a short interval (for example, 3 months) were also considered to progress within a longer time interval (for example, 6 months).

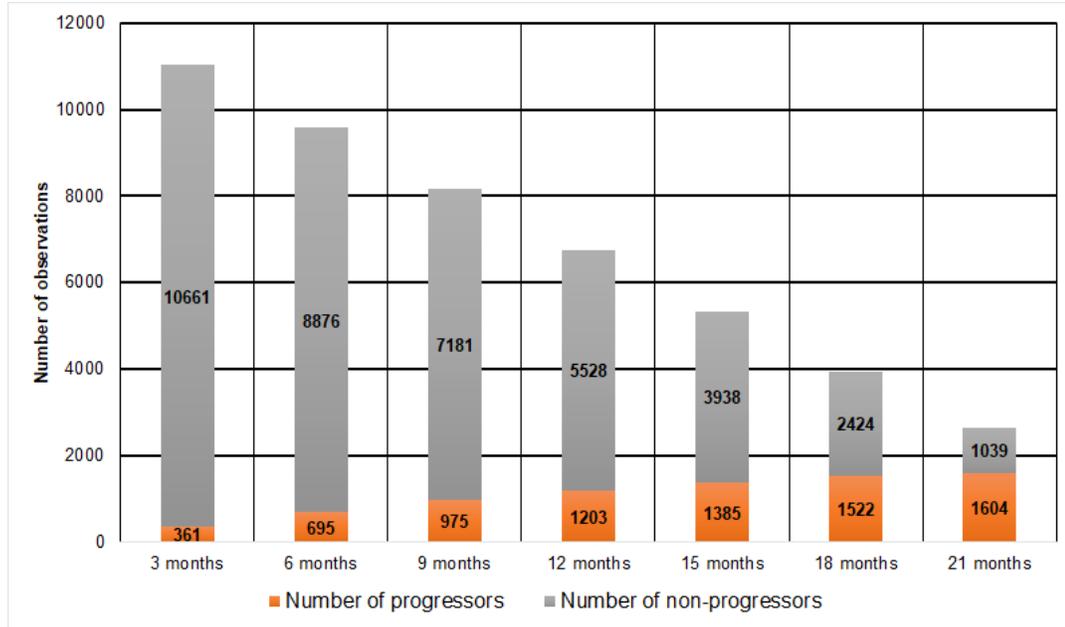

*Fig. 1. Study dataset distribution. Distribution of number of observations and outcomes considered for each of the time intervals analyzed in this study. The "progressors" label indicates a recorded progression event within the given time frame. The "non-progressors" label indicates certainty of not having an event within the given time frame.*

*Training and testing scheme*

Given the limited size of the study dataset, Random Forest and RNN models were evaluated in a 10-fold cross validation scheme, where the original sample (13,954 time points of 671 patients) was partitioned into 10 equal sized subsamples. The same proportion of progressing/non-progressing eyes was kept in each of the subsamples. We created a *patient level separation* while dividing the data in subsamples, where we train the model on the observations from 604 patients, and the observations from 67 patients are held out for validation. The *patient level separation* approach gives us more stringent evaluation of the model since we completely isolated the validation from the training set. Of the *10* subsamples, a single subsample of size roughly 1,395 time points is retained as the validation data for testing the model, and the remaining 12,559 subsamples are used as training data. The cross-validation process is then repeated *10* times, with each of the subsamples used exactly once as the validation data. The same 10-fold partitions were used to evaluate the RNN and Random forest model to provide a fair comparison between the two models.

As a primary metric, the accuracy of the prediction was evaluated by the Receiver Operating Characteristic (ROC) curve. ROC curve is a well-accepted method to show the trade-off between true-positive and false-positive prediction rates, where the models produce a probability scores for test samples, and present pairs of specificity and sensitivity values calculated at all possible threshold scores. These rates are irrespective of the actual positive/negative balance on the test set, and thus the ROC is insensitive to class distribution. The ROC curve also provides a single performance measure, the Area under the ROC curve (AUC) score, where AUC 1 represents a perfect prediction and 0.5 represents a random prediction.

*Predictive performance: Deep learning and traditional machine learning approaches*

We first evaluated the prediction models by considering the total number of observations (see *Overall performance*). For instance, if a patient has ten unique observations (visits at a given time point), 10 predictions are considered in the analysis whereas 2 predictions are considered for patients with two observations. Each visit is considered as an individual data point that includes features extracted from that particular visit and information from the previous visits of the same patient.

In order to understand the dependency of the model's performance with the number of historic visits considered, we also investigated the performance of Random Forest and Deep learning model (RNN) in a Visit-variant setting (see *Visit-variant performance*). We started the evaluation from 2 visits (1 current and 1 historic) to 16 visits (1 current and 15 historic) from the same eye and evaluated the performance of the models for 3 months up to 12 months in the AMD progression prediction task.

*Overall performance*

Side-by-side performance of the models in terms of AUC-ROC curve is shown in Fig. 2 using the same fold partitions. All the visits are considered as unique data point for this evaluation. We present the ROCs for all 10 folds as well as mean ROC (thick blue line). Being trained on the same OCT features, the RNN model performance is consistently better than the Random Forest model for all the prediction tasks with reasonably low variations between the folds. For instance, the mean AUC of RNN is 0.96 +/- 0.02 for prediction of AMD progression within 3 months from the visit while Random Forest scored 0.64 +/- 0.06 mean AUC. The high accuracy of the RNN model for both short and long-term progression suggests that it was able to integrate patient-specific demographics and imaging information by preserving long-term dependencies via the sequence-dependent temporal modeling. We observed a significant drop of RNN performance (p-value >0.05) for prediction of AMD progression within 12 months (0.77 mean AUC) compared to within 3 months, which improved for the 18 months. The performance drop could be due the fact that a fewer number of sequential visit data were available for model training, while characteristics of the progressor and non-progressor classes were quite close, which makes optimization of the prediction decision boundary the most difficult for the 12 months' time point. However, the performance improves at the 18 and 21 month time points, even with less data, since the training dataset is more balanced, while more diverse characteristics can be observed between long-term progressor and non-progressor classes.

**Random Forest prediction**                      **RNN prediction**

**3 months**

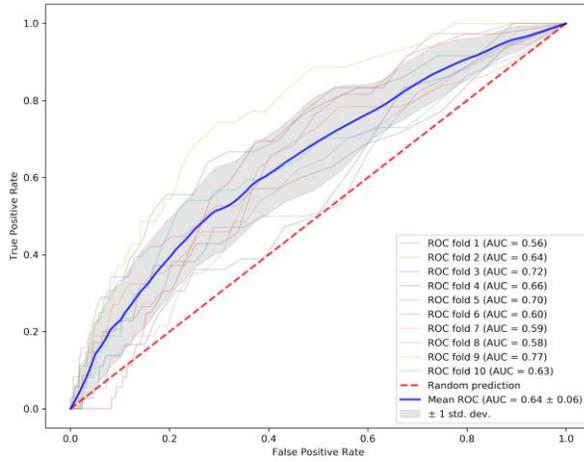
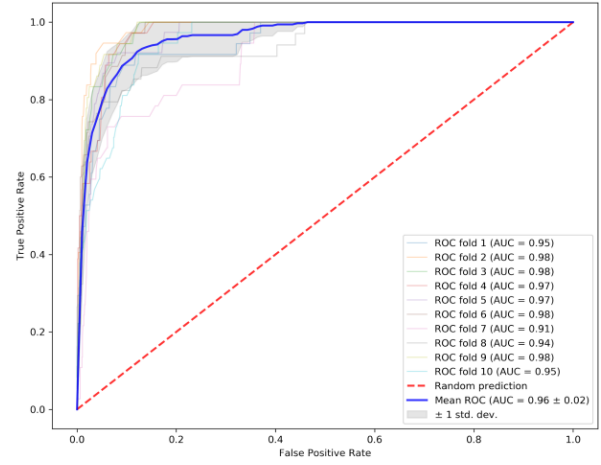

**6 months**

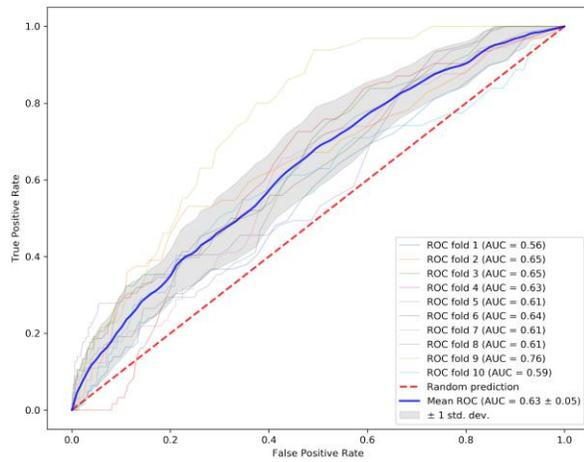
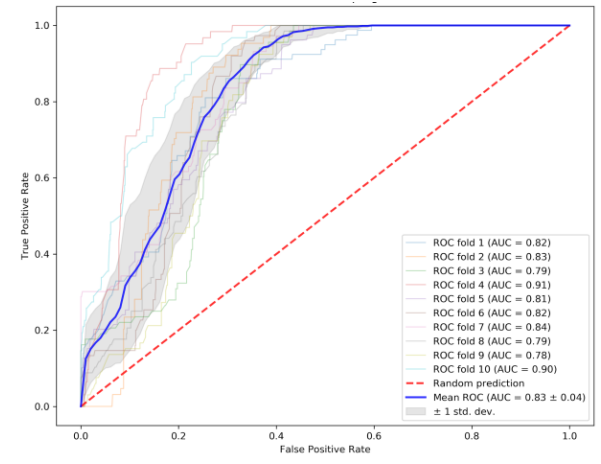

**9 months**

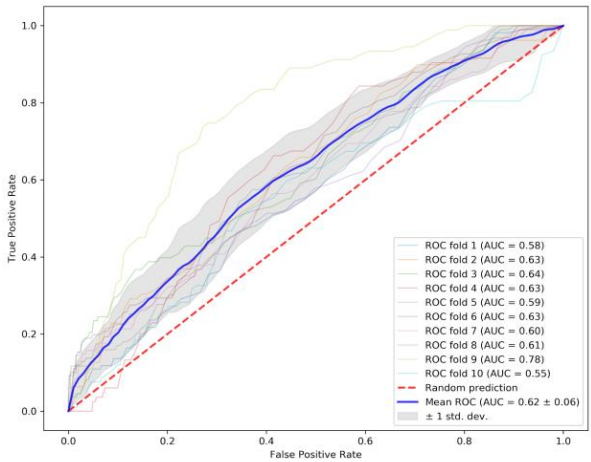
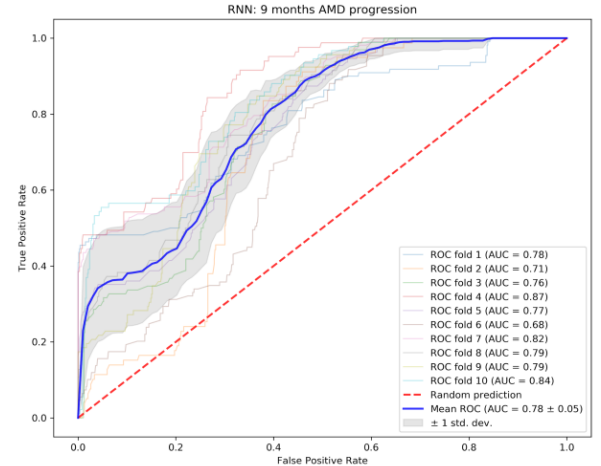

**12 months**

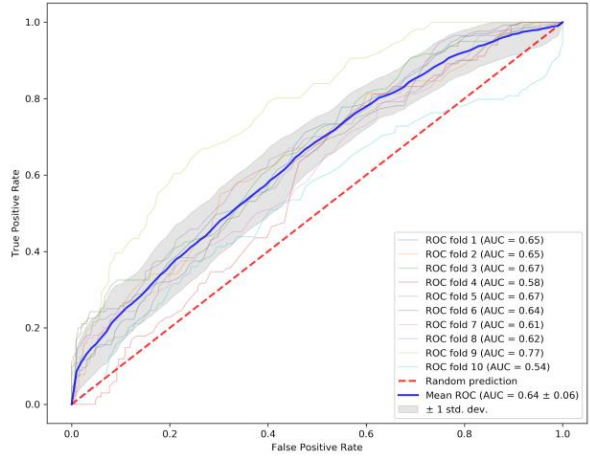
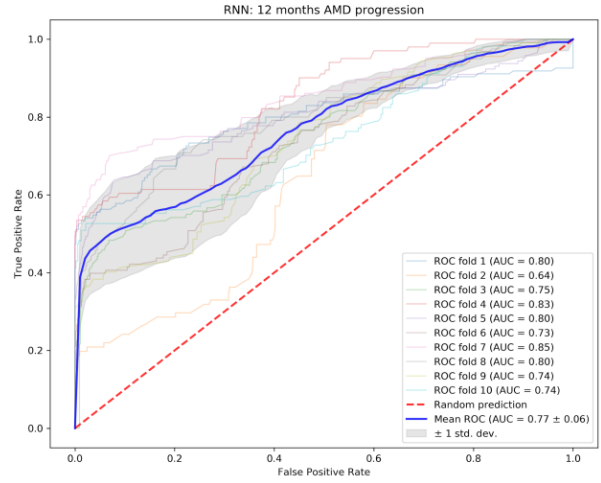

**15 months**

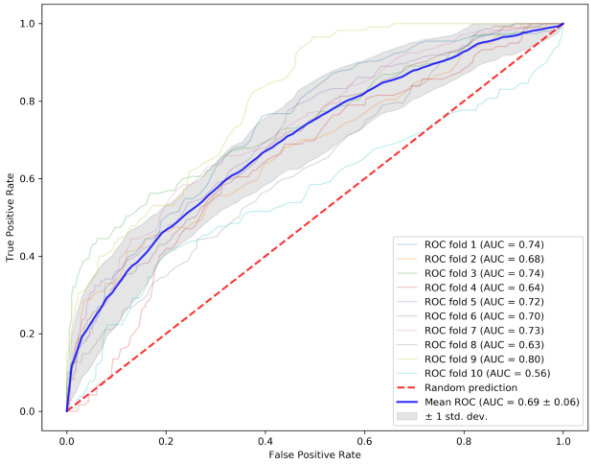
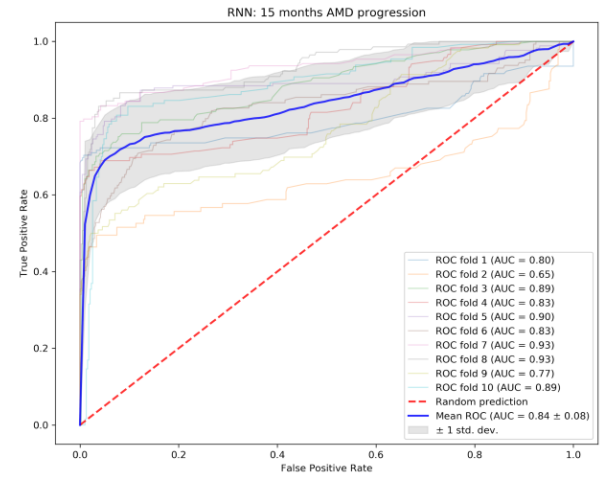

**18 months**

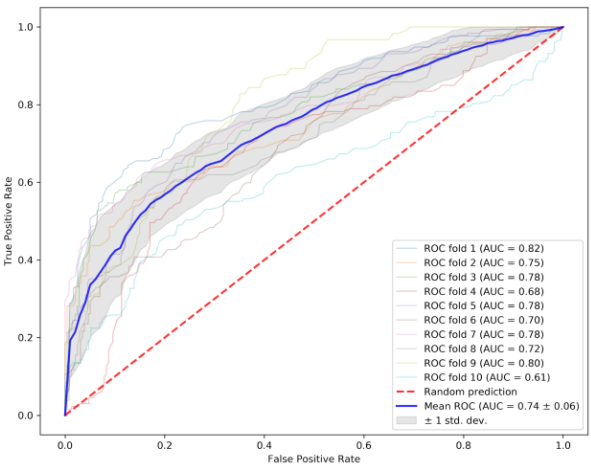
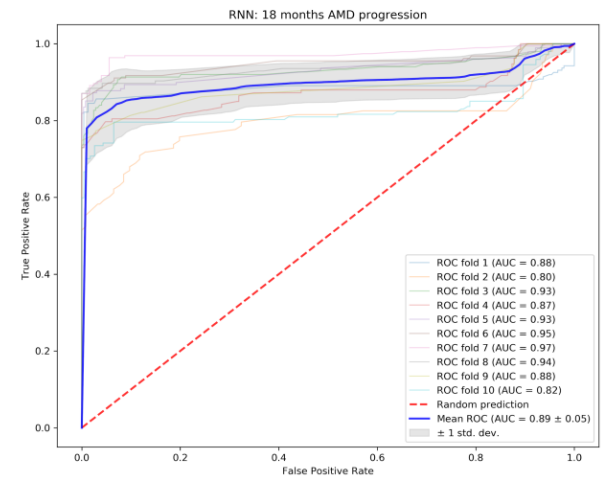

**21 months**

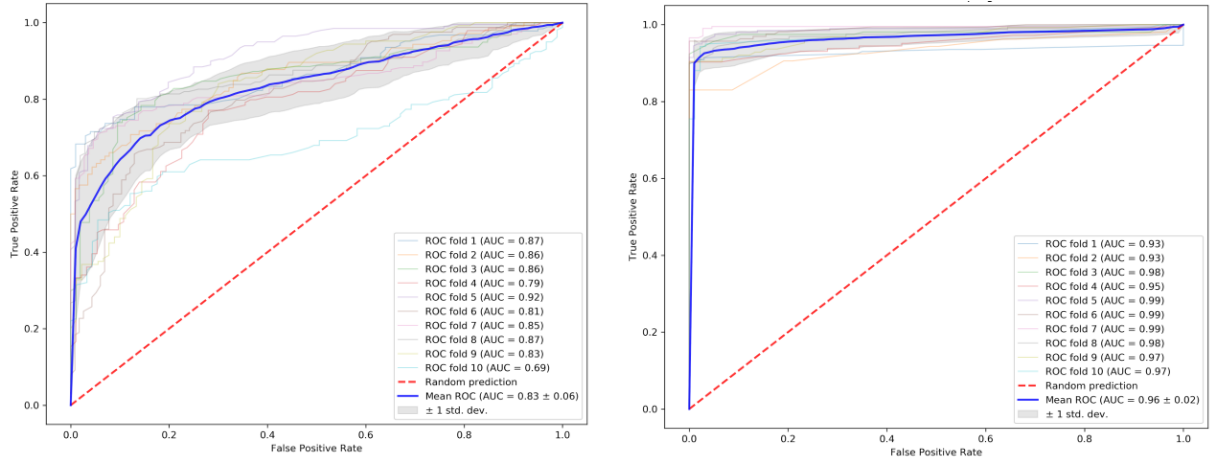

*Fig. 2: Side-by-side performance evaluation for Random Forest and RNN prediction model as ROC curve. Performance of Random Forest is presented in left and RNN model is presented in right.*

*Visit-variant performance*

Using the same training and testing separation, we validated both prediction models in a visit variant setting and show the AUC-ROC values in Fig. 3, where x-axis shows the number of visits and y-axis shows the AUC-ROC values. We present the performance for even number of visits from 2 up to 16. The reported performance shows that the Random Forest model's performance is not varying much with the number of historic visits while RNN prediction performance improves by increasing the number of historic visits. For instance, RNN performance increased from 0.64 up to 0.8 AUC-ROC for short-term progression prediction (within 3 months) by increasing the number of visits from 2 to 6, and from 0.68 up to 0.96 for long-term progression prediction (within 12 months) by increasing the number of visits from 2 to 16 visits.

      The improvement in RNN performance is likely due to the fact that the model understands the patient's current eye status better based on its understanding of previous visits by capturing long-term dependencies in visit sequences and, providing more historic visits, enhances the model's insight about the patient. Given this, the RNN prediction model is expected to perform better when a patient has more follow-up visits. However, with only 2 visits, the RNN model performs better than the Random Forest model for all the prediction tasks, which shows that even with little historic knowledge about the patient, the proposed RNN model can predict the progression of AMD better than the traditional model. This has potential clinical impact, since patients may not have data from many prior visits available.

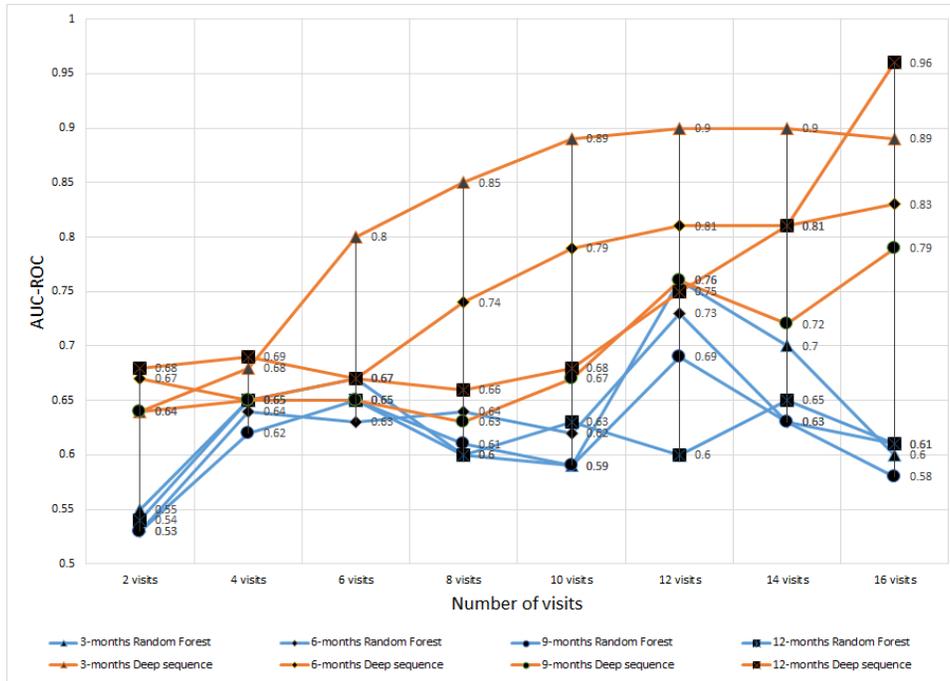

*Fig. 3. Visit-variant performance evaluation for Random Forest and RNN prediction model:* *blue colored lines shows the Random Forest performance and orange colored lines shows the RNN performance.*

**Discussion**

In this study, we proposed a sequence-dependent hybrid prediction model that, given a sequence of patient visits, can efficiently predict the probability of an exudative or CNV event from early and intermediate AMD (AMD progression) at different timepoints (within 3 months up to 21 months in 3 months intervals). The study showed a way of integrating advanced radiomics methods[17] with sequential deep learning for analyzing longitudinal clinical visit data to make prediction about disease progression. The engineered OCT features extracted by our radiomics pipeline allowed the proposed RNN model to focus only on the relevant predictors (e.g. druse volume, area) rather than raw pixel data, and discover complex relationships within sequence of visits by applying a long-term dependency learning between the predictors. The deep learning model was compared against a traditional random forest model that used the same available data to make a prediction based on a combination of demographics and quantitative image features. Both models were trained and evaluated in a cross-validation setting on 13,954 observations from 671 fellow eyes of patients considered in the HARBOR study.

    Being trained on the delta features (difference between current and historic feature values), the Random Forest model scored 0.66+/- 0.07 AUC for short-term (within 3 months) AMD progression prediction, while scoring 0.83+/-0.03 AUC for long-term (within 21 months) prediction. It can be observed that prediction for longer time intervals was more accurate than for shorter time intervals. These results are superior to those obtained in a previous study[16] using the same dataset as employed here. While previous study only aimed to predict the probability of progression at the end of the study considering the observations up to the fourth month of the

study, this translates to a prediction of AMD progression within 20 months and reported an AUC of 0.68 for predicting a CNV or exudation event. We hypothesize that the differences in performance may be a result of the different algorithm employed to automatically segment the OCT data in order to extract the imaging features, a different model used for the prediction (a Cox proportional hazards model instead of a Random Forest model) and/or the consideration of all possible observations for training the model in this work (use of LTSM units). This previous work also considered the genetic data included in the HARBOR dataset and modeled the prediction of GA appearance in a similar manner (0.8 AUC). Inclusion of genetic data as part of our models and expanding the ability of the models to predict GA appearance was not done here and will be a matter for future work.

The proposed RNN model out-performed the Random forest model by predicting short-term (within 3 months) AMD progression with 0.96+/-0.02 AUC and long-term (within 21 months) progression with 0.96+/-0.02 AUC. The high prediction accuracy suggests that the proposed model learns the complex longitudinal relation between the combination of demographic and visual factors and imaging features by analyzing a sequence of visits and can establish the correlation with both short-term and long-term AMD progressions. Visit-variant performance evaluation demonstrates an interesting insight that having many follow-up patient visits can boost the performance of the prediction model by learning patient-specific temporal trends. For instance, in the same setting, the performance upgraded from 0.64 to 0.89 AUC-ROC for short-term follow-up by increasing the number of visits from 2 to 16.

The American Academy of Ophthalmology recommends comprehensive eye examinations with OCT imaging every three months in patients more than 65 years of age with risk of AMD progression. Such longitudinal ophthalmic images are ideal for training a machine learning model to diagnose blinding eye diseases earlier and prevent disease progression by analyzing current and historic data. It has also been shown that demographic factors, like gender, age and smoking history, are confounding factors for AMD progression. Analysis of OCT images to predict progression of AMD has been recently applied to improve early diagnosis and prevention. de Sisternes et al [11], were among the early scientists to automatically extract drusen features from SD-OCT images to predict future AMD progression from an early or intermediate nonexudative stage (dry AMD) to an advanced exudative stage (wet AMD). However, previous studies did not consider quantitative image features, demographic, and visual factors from each sequential visit directly, but only considered the linear trend of the features from historic data (as done in the Random Forest model employed here). In the current study, we proposed a recurrent neural network which takes a long-sequence of visits as input for computing risk for AMD progression by remembering relevant information across long periods of time.

We consider the following three points as key limitations of our study.
i. **Trained and validated on the same clinical trial dataset -** The OCT data on eyes analyzed for the progression of AMD come from a single clinical trial, which is an artificial setting with bias selection criteria for the observations. Therefore, the dataset may not represent real-world patient data which may affect the generalizability of the model to an independent dataset. However, the high accuracy achieved by our RNN model in two different settings suggests that the model may able to predict AMD progression given the successful extraction of 21 imaging biomarkers. We adopted a standardized pipeline (segmentation and feature extraction) to handle the OCT images to ensure satisfying biomarker extraction performance. It is also important to note that the features used in the modeling were averaged through multiple scans collected at a single imaging session (an average of 2.02 scans per session). The acquisition of multiple scans may not be typical of clinical practice and may have added to the stability of the extracted features. The performance of the model considering only a single acquisition per imaging session in a separate dataset will be matter of study in upcoming future work.

ii. **Limited size of the observations -** For machine learning, an interesting and almost linear relationship can be observed between the amount of training data required and the complexity of the model, with the basic reasoning that the model should be large enough to capture relations in the data along with specifics of the targeted task. In our study, the RNN and random forest models were trained and validated on a limited number of observations - 13,954 time points of 671 patients, which can be considered as a limitation of the study. Thus, we restricted the learning parameters (neuron weights) of the RNN model to minimal count. However, the prospective data collection allowed us to represent significant samples for training. Additionally, we applied a data augmentation method to train the RNN model.

iii. **Limited explain-ability -** Given the complexity of the data (visit + imaging features) and the use of deep learning models, it is not trivial to explain the basis for predictions made by the models. For instance, in the RNN model, learned weights of the neurons can provide insights into the usefulness of the features, but it can also be biased by the combination of historic visit data. On the other hand, Random forest is popular for feature ranking, but these models may be challenged when data interpretation is important. With correlated features, strong features can end up with low scores and the method can be biased towards variables with many categories.

**Materials and Methods**

The OCT imaging/demographic data used was curated from a previously acquired longitudinal dataset that includes several OCT scans at different time points and a set of demographic features. The variables used in this study were previously described as indicators of possible disease progression[11] and were considered to train and evaluate two different models for the prediction of a CNV or exudative event from a non-neovascular AMD diagnosis: (i) A traditional model using Random Forest, and (ii) A deep learning predictive model using Recursive Neural Network (RNN).

*Demographic Factors*

A set of 5 demographic and visual features were considered in our analysis. These features were collected as part of the HARBOR dataset and included the following (as listed in Table 1): Age, gender, race, smoking status and visual acuity at baseline. Table 1 contains the average and standard deviation values for these features at the time of the first OCT observation available per patient.

***Table 1: List of demographics considered in our analysis.*** *Contains average and standard deviation (std) values at first available OCT observation overall and for fellow eyes with/without a progression event (progressors/non-progressors) during the study.*

| Demographic Feature | Description | All fellow eyes (N=671) | Progressors (N=149) | Non-progressors (N=522) |
|---|---|---|---|---|
| Age | Age of the patient in months at baseline mean (std) | 78.2 (8.3) | 79.5 (7.7) | 77.8 (8.4) |
| Gender | Patient gender: Male/Female % | 40.4 / 59.6% | 30.2 / 69.8% | 43.3 / 56.7% |
| Race | Patient Ethnicity: American or Alaska native / Asian / Black or African American / White / Native Hawaiian or Pacific Islander / Multiracial | 0.3 / 1.6 / 0.4 / 96.9 / 0.3 / 0 % | 0 / 0.7 / 0 / 98.7 / 0.7 / 0 % | 0.4 / 1.9 / 0.6 / 96.4 / 0.2 / 0 % |
| Smoking status | Smoking status: Non-smoker / Previous smoker / Current smoker | 41.0 / 48.4 / 10.6 % | 38.9 / 47.0 / 14.1 % | 41.6 / 48.8 / 9.6 % |
| Visual Acuity | Visual acuity at baseline of observation measured in LogMAR scale | 76.07 (13.07) | 76.91 (9.31) | 75.83 (13.96) |

*Extraction of imaging biomarkers*

A set of 21 imaging features describing presence, number, extent, density and relative reflectivity of drusen were extracted directly from each SD-OCT volume (Table 2). Drusen characteristics were computed automatically in a similar manner as studied as predictors of AMD progression in a previous publication[11]. Each OCT volume was processed using proprietary Cirrus Review Software (Carl Zeiss Meditec, Inc., Dublin, CA, United States) to automatically segment the location of the Retinal Pigment Epithelium (RPE) in the form of a surface and to generate a topographic map describing the regions of substantial RPE elevation (known as "Advanced RPE Analysis" in Cirrus Review Software). The location and extent of individual druse were also automatically segmented for each OCT volume using previously published methods [12], taking the segmented location of the RPE layer and an estimation of the inner segment/ outer segment junction (20 microns inner to the RPE locations) as inputs to automatically outline drusen within the volume. The result of this processing is a topographic map indicating RPE elevation with respect to a Bruch's membrane (BM) estimation (obtained from Cirrus Review Software) and the three dimensional segmentation of drusen locations for each considered OCT volume (obtained as described in [12]), where each individual druse is outlined. For a visualization of the segmentation results see Fig. 4.

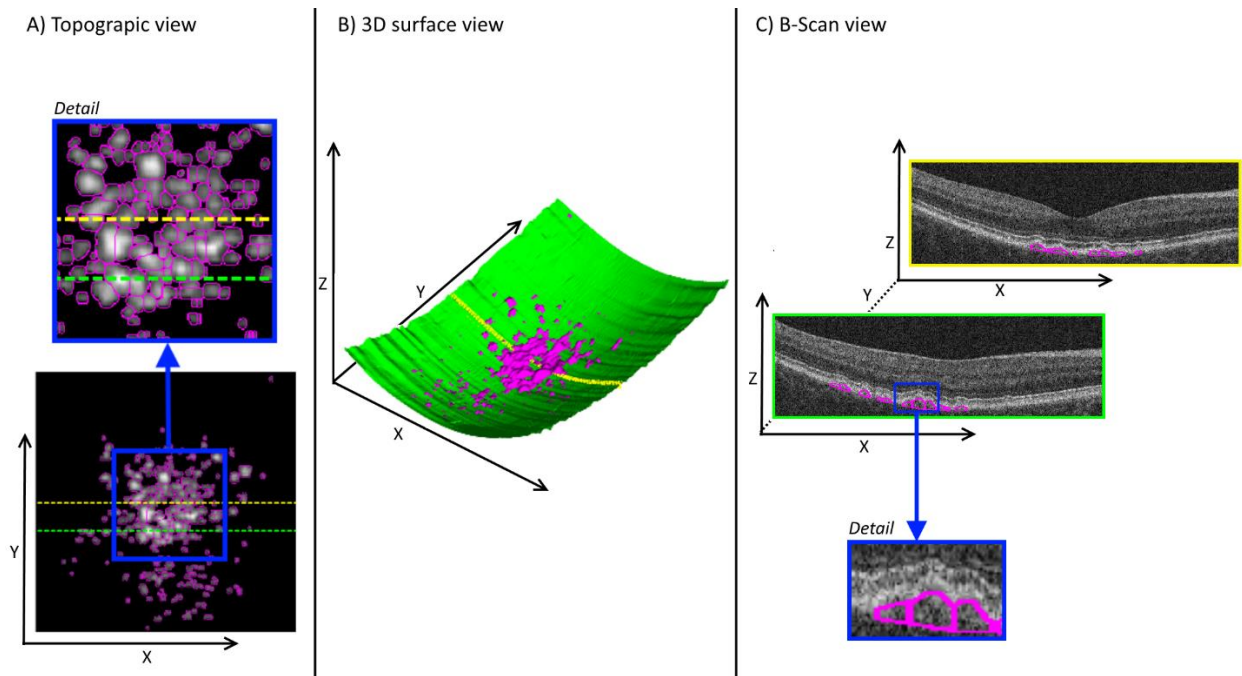

*Fig 4. Imaging biomarker extraction.* *Image shows a 3-D surface view of segmented drusen, with the estimation of BM surface indicated in green color and the detected druse regions identified in magenta. Druse identification in a volumetric manner allows the characterization of its volumetric properties. The image in the left shows a topographic view of a druse elevation map with individual drusen indicated in magenta (the blue square identifies a region shown in detail). The dotted green and yellow lines indicate B-scan locations shown in the right side. This topographic view allows the characterization of druse area, extent and density properties. The image in the right shows the individual druse segmentation in two example B-scans, with generated druse outlines shown in magenta. The blue square identifies a region shown in detail, where indications of drusen height, slope and reflectivity are shown. Consideration of the B-scan data allow the characterization of reflectivity properties inside druse regions.*

The topographic RPE elevation map and individual drusen delineations were processed to generate the set of features describing the number of independent drusen in SD-OCT volume, mean area per druse, total area occupied by drusen in topographic map, extent of topographic region affected by drusen (computed by generating the convex hull of the regions affected by individual drusen), drusen density in topographic map (computed as the ratio between the area occupied by drusen and the total region affected by drusen[11]), total drusen area within 3mm from the fovea center, and total druse area within 5mm from the fovea center.

The collection of B-scans within the SD-OCT volume and the corresponding three dimensional segmentation of drusen locations within the volume was processed to generate the following features: mean volume per druse, total volume occupied by drusen, maximum height with respect to BM observed for any druse in the collection of B-scans, average druse slope (computed as the gradient of drusen height with respect to BM), average relative reflectivity in the B-scans within regions delineated as drusen and standard deviation (std.) of the relative reflectivity in the B-scans within regions delineated as drusen. The relative reflectivity values were computed by normalizing the collection of B-scans so that the 95 percentile pixel value from the collection took a value of 1 and the 5 percentile took a value of 0. Since the RPE layer typically presents the highest reflectivity within the retinal layers in the scan and the vitreous region presents the lowest reflectivity, these values indicate a reflectivity ratio normalized between vitreous and RPE reflectivity as measured by the OCT system.

In addition, 6 druse features were also collected as reported automatically from the "Advanced RPE analysis" function in Cirrus Review software (Carl Zeiss Meditec, Inc., Dublin, CA, United States): area and volume within 3mm from the fovea center, area and volume within 5mm of fovea center, and total area and volume in scan field of view. These druse area and volume analyses differ from the previous olatlargerne in terms that the direct output from the Cirrus review software was considered here, whereas the individual druse segmentations computed as in [12] were considered in the previous analysis.

*Table 2:* *List of imaging features*

| Image Features | Description |
| --- | --- |
| Number of drusen | Number of individually separated drusen within OCT volume |
| Druse mean volume | Average volume occupied by each individual druse within the OCT volume (in $mm^3$/druse) |
| Druse total volume | Total volume occupied by all drusen within the OCT volume (in $mm^3$) |
| Druse mean area | Average area occupied by each individual druse within the OCT topographic map (in $mm^2$/druse) |
| Druse total area | Total area occupied by all individual druse within the OCT topographic map (in $mm^2$) |
| Extent of druse area | Total area affected by druse regions (convex hull of detected individual druse regions) within the OCT topographic map (in $mm^2$) |
| Druse density | Density of drusen in affected regions (Feature#10/Feature#11) within the OCT topographic map |
| Maximum druse height | Maximum height of drusen with respect Burch's membrane observed in collection of OCT B-scans (in mm) |
| Avg. druse slope | Average drusen slope (gradient of drusen height) within the OCT volume |
| Avg. druse reflectivity | Average value of normalized pixel intensity (values 0-1) inside drusen regions observed in collection of OCT B-scans |
| Std. druse reflectivity | Standard deviation of normalized pixel intensity (values 0-1) inside drusen regions observed in collection of OCT B-scans |
| Druse area 3mm | Area occupied by the all the individual druse regions in the OCT topographic map within 3mm from the fovea center (in $mm^2$) |
| Druse area 5mm | Area occupied by the all the individual druse regions in the OCT topographic map within 5mm from the fovea center (in $mm^2$) |
| Druse volume 3mm | Volume occupied by the all the individual druse regions in the OCT volume within 3mm from the fovea center (in $mm^3$) |
| Druse volume 5mm | Volume occupied by the all the individual druse regions in the OCT volume within 5mm from the fovea center (in $mm^3$) |
| Druse total area (Cirrus) | Area occupied by the all the individual druse regions in the OCT topographic map as provided by Cirrus review software (in $mm^2$) |
| Druse area 3mm (Cirrus) | Area occupied by the all the individual druse regions within 3mm from the fovea center as provided by Cirrus review software (in $mm^2$) |
| Druse area 5mm (Cirrus) | Area occupied by the all the individual druse regions within 5mm from the fovea center as provided by Cirrus review software (in $mm^2$) |
| Druse total volume (Cirrus) | Volume occupied by the all the individual druse regions within the OCT volume as provided by Cirrus review software (in $mm^3$) |
| Druse volume 3mm (Cirrus) | Volume occupied by the all the individual druse regions within 3mm from the fovea center as provided by Cirrus review software (in $mm^3$) |
| Druse volume 5mm (Cirrus) | Volume occupied by the all the individual druse regions within 5mm from the fovea center as provided by Cirrus review software (in $mm^3$) |

*Traditional predictive model: Random Forest*

We employed a Random Forest model to describe the probability of AMD progression within a given time frame. Random Forest is a non-linear ensemble learning method that operates by constructing a multitude of decision trees based on the analyzed features to predict an outcome[19]. Bootstrap-aggregated (bagged) decision trees combine the results of these many decision trees, which reduces the effects of overfitting and improves generalization. We constructed 7 independent models to characterize conversion probability of a given observation within 7 different time frames: 3, 6, 9, 12, 15, 18 and 21 months from the observation time. Each of the trained models employed an ensemble of 200 decision trees, grown using bootstrap samples of the data. Gender, race and smoking status were treated as categorical features, while the rest of demographic features and imaging features were treated as numerical features. An additional 21 imaging features were also included characterizing the relative rate of change of each of the imaging characteristics along time up to a considered observation. The rate of change was computed by fitting a linear function along time for each independent imaging feature listed in Table 2, considering the collected previous observations from the same eye up to and including the considered observation and reporting the inclination (first coefficient) of the linear fit.

Each of the models used a regression ensemble[20] where the outcome was the progression (1) or non-progression (0) to wet AMD within a given time frame. After the 7 independent Random Forest models were trained, the output of the model for an observation not seen in the training data was a set of 7 scores ranging from 0 to 1, each describing the probability of progression from within 3 months up to within 21 months in 3 month intervals, respectively.

*Deep learning model: Recursive Neural Network (RNN) prediction model*

We designed a many-to-many RNN model (see Fig. 5) using two-layer one-directional stacked stateful Long short-term memory (LSTM) units[21] to predict progression from dry to wet AMD across the sequence of clinical visits. We chose LSTM because it is relatively insensitive to gap length compared to alternatives such as RNNs and hidden Markov models[22]. The long-term memory allows slow weight updates during training and encodes general information about the whole temporal visit sequence, while short-term memory has an ephemeral activation, passing immediate state between successive nodes and resetting itself if a fatal condition is encountered. The LSTM includes memory about prior observations (patient visits) and thus accounts for longitudinal changes in the patient data. The model takes as input a series of a combined feature matrix ($Feature\ vector@O_n$ in Fig. 5) –consisting of the described demographic and quantitative imaging features, ordered according to timestamp of observation where $O_n$ stands for observation at n, and predicts probability of AMD progression. Our objective is to predict AMD progression at multiple future timepoints, starting from short-term progression (3, 6, 9 and 12 months), up to long-term progression (15, 18 and 21 months); thus, we formulate a set of single time-point prediction models. Each RNN model predicts the survival rate at only one time point($T_i$), and each of the seven time points was analyzed by the same architectural model but trained separately.

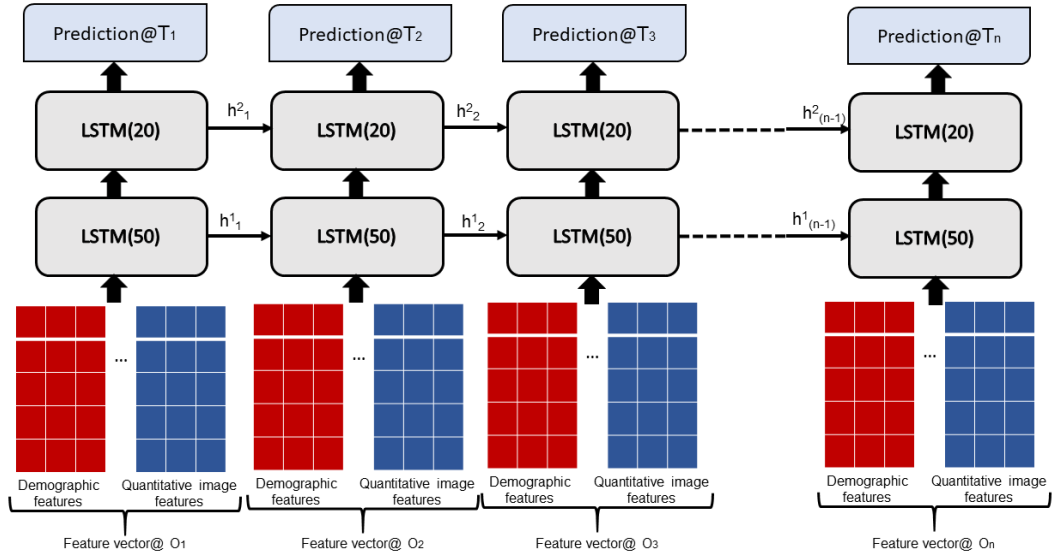

***Fig. 5*** *Many-to-many LTSM model for predicting progression of AMD*

For each patient (with id $i$), the sequence of feature vector is modeled as a series: $X_i = \{x_i^1, x_i^2, \ldots\ldots, x_i^n\}$, where each input data observation $x_i^t \in R^D$ is a real-valued vector representation of the demographic and quantitative imaging features at the time of observation $t$ and n is the total number of observations for the patient. Continuous imaging features are mapped to float values, and categorical features are also embedded into numeric values. All the features are concatenated into one feature vector and passed into the stacked neural network. The targeted AMD progression sequence for month $k$ of the patient with id $i$ is modeled as: $Y_i = \{y_i^1, y_i^2, \ldots\ldots, y_i^n\}$, where $y_i^t \epsilon \{0,1\}$ is a categorical variable that represents whether the patient will have a progression event within month k starting from the observation timepoint. Single directional stacked LSTM units are modeled to encode sequence-dependency between the longitudinal visit and predict a probability of progression for each time point, following the principle that at the time point $t$ the model does not have access to the future feature information $x_i^{t+1}$ but can access data from the current and all the historic time points: $\{x_i^1, x_i^2, \ldots\ldots, x_i^{t-1}, x_i^t\}$. A sequence of patient with id $i$ with $n$ number of visit/observation is defined as: $S_i^n = \{(x_i^1, y_i^1), (x_i^2, y_i^2), \ldots, (x_i^n, y_i^n)\}$

Data augmentation strategy for temporal visit series: Being a supervised machine learning approach, our proposed model is limited by the number of available training examples. In this study, we explore a simple data augmentation technique (only on the training set) which not only boosts the number of training sequences, but also handles varying number of visits for different patients. Given a sequence $S_i^n$ of $n$ length, we created $n$ augmented sequences $AugS_i^l$ of length $l$ by incrementally adding data points at the end of the sequence and padding the rest, where the first sequence contains 1 data point $(x_i^1, y_i^1)$ and $(l-1)$ post-padding points: $P = (z, y_z)$, z is a zero vector of dimension $D$ and $y_z = 2$. Similarly, the second sequence contains 2 data points, third sequence contains 3 data points, and so on (see Fig. 6).

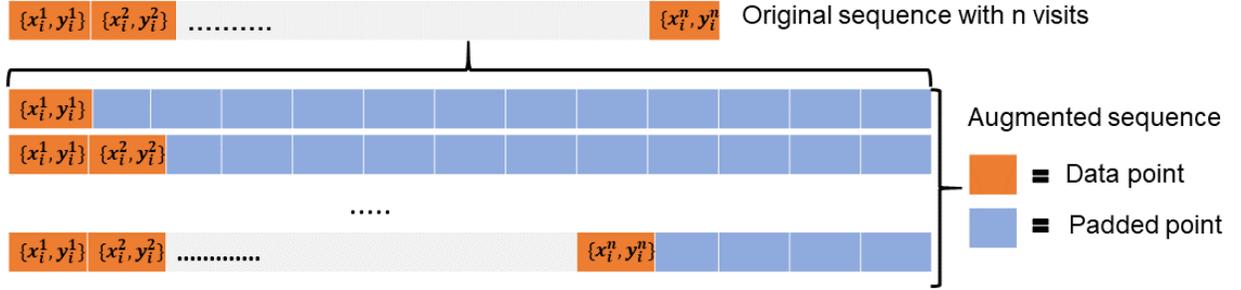

***Fig. 6*** *Augmentation of the visit sequence for training the neural network model*

In the stacked RNN layers, the first layer's one directional LSTM block receives the input $x^t$ and previous hidden state $h^{t-1}$ and passes the current hidden state $h^t$ to the successive LSTM blocks. The stacked layers enable more complex representation of our temporal series data, capturing information at different scales. The first layer's block also passes the hidden state and current H-dimensional cell state $c_t \epsilon R^H$ to the corresponding block in the upper layer. The second layer units are modeled to maintain the recurrent connections in multiple dimensions. The output estimate would be a vector of probabilities across three different labels: $L = \{0, 1, y_z\}$, where 0 means no progression, and 1 stands for progression. The output at each timepoint $t$ is modeled as: $\widehat{y^t} = softmax(L. h^t)$, where $\widehat{y^t} \in R^3$ is the predicted progression at time $t$ and $h^t$ the hidden state of the second layer LSTM. The three trainable parameters of each LSTM block are – (i) input-to-hidden weight matrix: $W_x \epsilon R^{4H \times D}$ (ii) hidden-to-hidden weight matrix: $W_h \epsilon R^{4H \times H}$, and (iii) bias vector: $b \epsilon R^{4H}$.

During the training phase, our model takes as input the augmented series $(AugS_i^l)$ for all the patients present in the training set, and optimizes the time distributed weighted cross entropy loss function:

$$l(Y, \hat{Y}) = -\frac{1}{l}\sum_{t=1}^{l}(y^t ln\widehat{y^t} + (1-y^t)ln(1-\widehat{y^t})).\lambda^k$$

where $y^t$ actual reference survival at $t$th time point in the sequence, $\widehat{y^t}$ represents the output of the neural network given the current sequence inputs, $\lambda^k \epsilon R^3$ and corresponds to the pre-defined weights of the three targeted labels $L$. We present a folded configuration in Fig.7 with a layer 1

LSTM block with 50 hidden neurons and layer 2 block with 20 neurons, where the selection is a tradeoff between the input data dimension and the memory requirement for training.

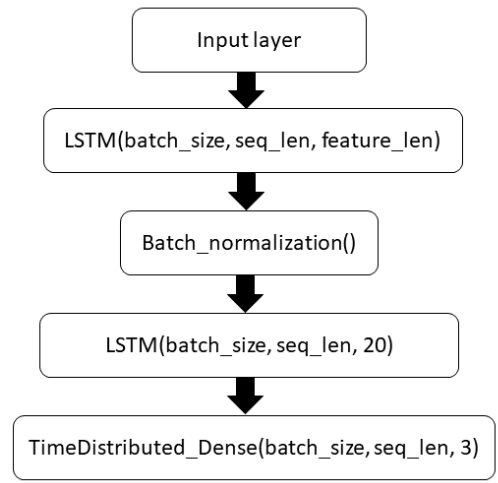

*Fig. 7* Architecture of the LSTM model (folded)

**Acknowledgments: Funding**: This project was supported by a grant from Genentech Inc. **Author contributions**: IB, LS, and DL conceived the project and led the studies. AO provided the study data. LS and MD extracted the image features. IB and LS. carried out the design of the machine learning models and performed the experiments. IB and LS wrote the manuscript, and all authors edited the manuscript. **Competing interests:** The authors declare that they have no competing interests. **Data and materials availability**: The prediction models can be obtained through an MTA.